\documentclass[a4paper,12pt]{article}
\usepackage{amssymb,amsmath,amsthm}
\usepackage{cite}
\usepackage{hyperref,subcaption}
\usepackage{authblk}
\usepackage{tikz,tikz-cd,pgfplots,tkz-euclide,circuitikz}
\pgfplotsset{compat=1.17}
\usetikzlibrary{arrows,positioning,calc,fadings,decorations.pathreplacing,decorations,decorations.text,decorations.markings,quotes,angles,fit,spy}
\usepgfplotslibrary{patchplots,colormaps}
\newcommand\C{\mathbf{C}}

\newcommand\Cs{\mathbf{C}^{\star}}
\def\F{\mathcal{F}}
\def\tF{\tilde{\mathcal{F}}}
\def\G{\mathcal{G}}
\newcommand\cp{{\mathbf{CP}}}

\newcommand\Ox[2]{{\mathcal{O}}_{#1}({#2})}
\newcommand\braket[2]{\langle#1|\,#2\rangle}
\def\kahler{\mathcal{K}}
\def\tkahler{\tilde{\mathcal{K}}}
\def\tz{\hat{z}}
\def\pa{\partial}
\newcommand\eq[1]{(\ref{#1})}
\newcommand\bra[1]{\langle#1|}
\newcommand\ket[1]{|#1\rangle}
\theoremstyle{definition}
\newtheorem{example}{Example}
\title{On entropy and complexity of coherent states and K\"ahler geometry}
\author{Koushik Ray\thanks{email: koushik@iacs.res.in}}
\affil{\normalsize Indian Association for the Cultivation of Science,
\authorcr Calcutta 700 032. India.}
\date{}
\begin{document}
\maketitle
\begin{abstract}
\noindent 
Consanguinity of entropy and complexity is pointed out through the example
of coherent states of the group $SL(d+1,\C)$. Both are obtained from the 
K\"ahler potential of the underlying geometry of the sphere 
corresponding to the Fubini-Study metric. Entropy
is shown to be equal to the K\"ahler potential written in terms of 
dual symplectic variables as the Guillemin potential for toric manifolds. 
The logarithm of complexity relating two 
states is shown to be equal to Calabi's diastasis function.
Optimality of the Fubini-Study metric is indicated by considering its 
deformation. 
\end{abstract}
\setcounter{page}{1}
\thispagestyle{empty}
\section{Introduction}
Entropy and complexity are two important notions in statistical theories. 
Entropy is the enumeration of different ways of organizing states 
of  a system, often looked upon as lack of order. 
Complexity is a measure of difficulty in evolving from one state to another. 
Various quantitative definitions of both the notions have been 
proposed and studied in various contexts. Relation between these 
has also been sought in different areas \cite{Li,KKS,MV,Modis,GV,Shan}.
They have appeared simultaneously in the context 
of holographic gauge-gravity duality in recent times. 
Inadequacy of the holographic 
entanglement entropy in probing black holes beyond the
event horizon is believed to be supplemented by the gravitational dual 
of circuit complexity 
\cite{nielsen2006quantum,nielsen2005geometric,dowling2008geometry}
in the boundary theory 
\cite{suss1,suss2,suss4,moshen}.
Various geometric aspects of entropy are known \cite{BZ}.
Another geometric interpretation of complexity may be useful. 

Interrelations between entropy and complexity or other information 
functions in full generality is difficult to conceive. 
Studying it within the scope of examples in different
contexts appears to be a more pragmatic approach. 
In here, we discuss these concepts and their K\"ahler geometric 
provenance in the simple example of coherent states of the $SU(d+1)$
group, or rather, its complexification, $SL(d+1,\C)$.

Let us present a pr\'ecis of the identifications at the outset.
Coherent states of $SL(d+1,\C)$ are vectors in the highest weight 
module of the group. 
Spin $j$ coherent states are given by polynomials of degree $2j$ in 
$d$ complex variables
$(z_1,z_2,\ldots,z_d)$. Geometrically,
these are looked upon as global sections of a degree $2j$
line bundle on the complex projective space $\cp^d$, generalising the
so-called Bloch sphere for $d=1$. Defining their inner product suitably,
taking into account this geometric picture, allows us to normalize the
coherent states \eq{xi:fin}.
A probability distribution can be defined for the normalized
coherent states, leading to entropy of von Neumann or
Shannon type \eq{Weh}.

The projective space $\cp^d$ is a complex K\"ahler manifold with constant 
scalar curvature. A symplectic structure and thence a symplectic potential 
can be associated to it. We show that the  entropy of the spin $\tfrac{1}{2}$ 
coherent states associated to 
the degree $1$ line bundle $\Ox{\cp^d}{1}$ 
is the symplectic Guillemin potential \eq{G:pnk}. 

Circuit complexity is related to the geodesic distance between points in 
an appropriately defined metric in a manifold to which the states belong.
It has been studied for coherent
states of scalar field theories and fermions \cite{rcF,chetan,ghmr,
Yang,chapman2018toward}. We point out that circuit complexity 
of these coherent states can be thought of as Calabi's diastasis
function \eq{calabi:d}, which is computed directly 
from the Fubini-Study K\"ahler potential and equal to the logarithm of the 
overlap of two coherent states through the 
inner product alluded to above \cite{ber1,SV}. 
This shows that entropy and
complexity for these coherent states are consanguine,  
derived from the K\"ahler potential of
the underlying space. We briefly discuss a certain reasonable 
deformation \cite{calabi} of the Fubini-Study metric preserving the K\"ahler 
structure of the projective space \eq{K:deform}.
We observe that the identification of the diastasis function and
the circuit complexity fails for such deformed metrics. The Fubini-Study form
appears to be the optimally costly notion of complexity.
\section{Coherent states of $SL(d+1,\C)$}
Let us start by briefly recalling the construction of coherent states of
$SL(d+1,\C)$ \cite{perel,GS}. 
Let us define the integer $k=2j$, where $j$ may be half-integral.  
Let  $m=(m_1,m_2,\ldots,m_d)$ denote a $d$-tuple of integers and define
\begin{equation}
|m| = m_1+m_2+\cdots + m_d.
\end{equation}
A  basis of the highest weight module
with spin $j$ is given by the monomials
\begin{equation}
\begin{split}
\label{Vj}
\chi_k^m(z) &=z^{m} \\
& := z_1^{m_1} z_2^{m_2} \cdots z_d^{m_d},
\end{split}
\end{equation}
such that $|m|\leqslant k$. 
An inner product on the space of monomials is required
in order to interpret these as normalized quantum states. 
It is obtained by recognizing  the monomials $\chi_k^m(z)$ as 
global sections $H^0(\cp^d,\Ox{\cp^d}{k})$ of the degree $k$ line bundle
$\Ox{\cp^d}{k}$ over the complex projective space $\cp^d$,
$z$ denoting the affine coordinate of $\cp^d$.
Let  $(\tz_0,\tz_1,\tz_2,\ldots,\tz_d)$ be 
any non-zero point in $\C^{d+1}$. 
The complex projective space is obtained
as the quotient $\cp^{d} = \big(\C^{d+1}\setminus\{0\}\big)/\Cs$
by identifying points under dilation, that is, through
the equivalence relation
\begin{equation}
(\tz_0,\tz_1,\tz_2,\ldots,\tz_d) \sim 
	(\lambda \tz_0,\lambda \tz_1, \lambda\tz_2,\ldots,\lambda\tz_d),
\quad\lambda\in\C^{\star},
\end{equation}
where $\C^{\star}=\C\setminus\{0\}$ is the multiplicative group
of non-zero complex numbers, called the algebraic torus. 
In the coordinate chart $\tz_0\neq 0$, the projective space is given by a copy 
of $\C^d$ with coordinate
\begin{equation}
\label{cpd}
	z= (z_1,z_2,\ldots,z_d), \quad z_i = \tz_i/\tz_0, 
\end{equation}
invariant under the action of $\C^{\star}$.
The global sections of $\Ox{\cp^d}{k}$ are polynomials in 
the $d$ variables $z_1,z_2,\ldots,z_d$, with degree at most $k$. 
The natural K\"ahler form on $\cp^{d}$ is the Fubini-Study form 
given by
\begin{equation}
	\label{omOm}
\omega=\tfrac{i}{2\pi}\pa\bar{\pa}\kahler(z,\bar{z}) dz\wedge d\bar{z},
\end{equation}
in the usual notation, with the K\"ahler potential 
\begin{equation}
	\label{omg:def}
	\kahler(z,\bar{z}) =\ln(1+\Vert z\Vert^2),
\end{equation}
where we use the shorthand
\begin{equation}
\label{modz}
\Vert z\Vert^2=|z_1|^2+|z_2|^2+\cdots+|z_{d}|^2.
\end{equation}
The Fubini-Study metric obtained from it is  given as
\begin{equation}
\label{FS}
g_{i\bar{j}} = \pa_i\bar{\pa}_j\kahler(z,\bar{z}).
\end{equation}
The first Chern class of the bundle 
$\Ox{\cp^d}{k}=\Ox{\cp^d}{1}^{\otimes k}$ is 
given in terms of the K\"ahler form, 
$c_1\big(\Ox{\cp^d}{k}\big)=k\omega$.
Let us note that adding an arbitrary holomorphic or 
anti-holomorphic term to the 
potential does not alter the K\"ahler form $\omega$.

The space of sections $H^0(\cp^d,\Ox{\cp^d}{k})$ is bestowed with an inner 
product \cite{stz03}. The inner product of two polynomials ${f}$ and ${g}$ 
of degree $k$ each is given by 
\begin{equation}
	\label{norm:a}
	\langle {f},{g}\rangle = \binom{k+d}{d}
	\int_{\C^d} \frac{{f}(\bar{z}){g}(z)}{(1+\Vert z\Vert^2)^k}
	\omega.
\end{equation}
The inner product of two monomials 
$\chi_k^m(z), \chi_k^{m'}(z)\in H^0(\cp^{d},\Ox{\cp^{d}}{k})$, defined in
\eq{Vj}, is then given by   
\begin{equation}
	\label{inn:z}
	\langle \chi_k^m(z),\chi_k^{m'}(z)\rangle = 
        \frac{(k+d)!}{k!}
	\int_{\C^{d}} \frac{\chi_k^m(z){\chi_k^{m'}(\bar{z})}}{
(1+\Vert z\Vert^2)^{k+d+1}}
	dz_1d\bar{z}_1dz_2d\bar{z}_2\cdots dz_{d}d\bar{z}_{d}.
\end{equation}
The monomials are orthonormal with respect to this inner product. The norm 
of a monomial induced by this inner product is 
\begin{equation}
	\Vert\chi^m\Vert^2 
=\frac{1}{\binom{k}{m_1,m_2,\cdots,m_d}},
\end{equation}
where the denominator is a multinomial, namely,
\begin{equation}
\binom{k}{m_1,m_2,\cdots,m_d}= \frac{k!}{(k-|m|)!m_1!\cdots m_d!},
\end{equation}
and repeated use of the integral 
\begin{equation}
	\int_0^{\infty}\frac{x^a}{(1+x)^{b}}dx=
	\frac{\Gamma(1+a)\Gamma(b-a-1)}{\Gamma(b)}
\end{equation}
has been made in order to evaluate the integrations in \eq{inn:z}.

The coherent states of $SL(d+1,\C)$ with spin $j$ 
are expressed in terms of the orthonormal monomials \eq{Vj} as
\begin{equation}
\label{z:eq}
\ket{z}  = \sum_{\substack{m\\|m|\leqslant k}} \psi_{j,m}(z)\ket{j,m},
\end{equation} 
where $k=2j$ and  $\ket{j,m}$ 
denotes the basis states of the highest weight module, and 
\begin{equation}
\begin{split}
	\label{xi:fink}
	\psi_{k,m}(z) &= 
\frac{1}{(1+\Vert z\Vert^2)^{k/2}} \frac{\chi_{k,m}(z)}{\Vert\chi_{k,m}\Vert^2}
\\
&=\frac{1}{(1+\Vert z\Vert^2)^{k/2}}
        \sqrt{\binom{k}{m_1,m_2,\cdots,m_d}}
	z_1^{m_1}z_2^{m_2}\cdots z_{d}^{m_{d}}
\end{split}
\end{equation} 
denotes the wavefunction. 
We have incorporated the contribution from the fiber in the definition of the 
state so that the subsequent integrals are performed over the K\"ahler volume
only \cite{perel}. The Hilbert space inner product is then defined in terms of
\eq{inn:z} as
\begin{equation}
	\label{xixi}
	\braket{z'}{z}	= \sum_{\substack{m\\ |m|\leqslant k}}
{\psi}_{k,m}(\bar{z}')\psi_{k,m}(z)
=\left(\frac{1+\bar{z}'\cdot z}{%
			\sqrt{1+\Vert z'\Vert^2}\sqrt{1+\Vert z\Vert^2}
			}\right)^k,
\end{equation}
where we define
\begin{equation}
\bar{z}'\cdot z = \bar{z}'_1z_1+\bar{z}'_2z_2+\cdots+\bar{z}'_dz_d,
\end{equation}
and 
$\bra{z}$ denotes the Hermitian conjugate of $\ket{z}$ 
with respect to this inner product.
The integral of $|\braket{z}{z}|^2$ over $\cp^d$ with the Fubini-Study metric
is unity, fixing the normalization of the states. 

Let us restrict attention to $k=2j=1$. 
Clearly, the basis of monomials \eq{Vj} is indexed by the integer 
partitions of $k$ of length $d$, the number of such monomials being 
$\binom{k+d}{d}$, used in \eq{norm:a}. 
In the special case of $k=1$ there are $(d+1)$ monomials,
which we denote suppressing $k$ from the notation as
\begin{equation}
\label{chi:m}
	\chi^{0} = 1,\
	\chi^{1} = z_1, \ldots,\ \chi^{d} = z_d,
\end{equation}
by a slight abuse of notation.
A coherent state in this basis is given by
\eq{z:eq}, with $j=1/2$ and 
\begin{equation}
	\label{xi:fin}
	\psi_i(z) = \frac{\chi^i}{\sqrt{1+\Vert z\Vert^2}}.
\end{equation} 

A notion of entropy has been associated to coherent states. From \eq{xi:fin},
we have the probabilities 
\begin{equation}
	P_i = |\psi_i|^2, \quad i=0,1,\ldots,d,
\end{equation}
summing up to unity.
Shannon entropy is then defined as \cite{weh78}  
\begin{equation} 
\label{Weh}
	\begin{split}
		S &= -\sum_{i=0}^d P_i\ln P_i\\
		& = \ln (1+\Vert z\Vert^2) 
-\sum_{i=1}^d\frac{|z_i|^2\ln |z_i|^2}{1+\Vert z\Vert^2}.
	\end{split}
\end{equation} 
\section{Entropy as Guillemin potential}
\label{E2G}
Let us now introduce the symplectic potential \cite{gui,abreu,calder}. 
This has been used to obtain Ricci flat metrics on orbifold moduli spaces of
D-branes \cite{kr1,kr2} as well as for ACG metrics \cite{ashwin}.
Expressing  the affine coordinates of $\cp^d$ introduced in \eq{cpd} as
\begin{equation}
	\label{zeta}
	z_i = e^{\eta_i/2} e^{i\theta_i},
\end{equation}
and defining its modulus as 
\begin{equation}
	\label{xeta}
	x_i=|z_i|^2 = e^{\eta_i},
\end{equation}
the K\"ahler potential  \eq{omg:def}  becomes 
\begin{equation}
	\label{F:pn}
	\F(x) = \kahler(z,\bar{z})=\ln (1+\sum_{i=1}^d x),
\end{equation}
where we write the variables collectively as 
$x=(x_1,x_2,\cdots,x_d)$.
Invariance of the potential under the 
action of the maximal compact subgroup of the algebraic torus 
$\C^{\star}$, namely, the circle $S^1$ parametrized by $\theta_i$ is manifest
in this form.
The potential $\F$ can be obtained through the image under the moment map 
of the so-called Delzant polytope \cite{anna,kr1}. 
Let us define the dual variables
\begin{equation}
	\label{dual:y}
	y_i=\frac{\pa\F}{\pa\eta_i}.
\end{equation}
The Legendre transform of $\F$ with respect to $\eta$
is the Guillemin potential 
\begin{equation}
	\label{leg}
		\G =\sum_{i=1}^d\eta_iy_i-\F,
\end{equation}
which in the dual variables reads
\begin{equation}
\label{G:pnk}
	\G=\sum_{i=1}^d y_i\ln{y_i} +(1-{\sum_{i=1}^d y}) 
\ln (1-\sum_{i=1}^d {y}),
\end{equation}
where \eq{dual:y} is inverted to express $x$ in terms of $y$ as
\begin{equation}
	\label{yxeta}
	\begin{split}
		x_i=\frac{y_i}{1-\sum\limits_{i=1}^d y}.
	\end{split}
\end{equation}
The potential expressed in this form resembles the formula for Shannon
entropy with $y_i$ looked upon as probabilities. 
Indeed, expressing the entropy \eq{Weh} in terms of the dual variable $y$ using 
\eq{zeta}, \eq{xeta} and \eq{yxeta} we find 
\begin{equation}
	S = -\G.
\end{equation}
The Guillemin potential in this interpretation is also the entropy of number
of qubits \cite{Shan,belhaj}.
The qubits are then in one-to-one correspondence with the Cartier divisors
of the toric variety $\cp^d$.
\section{Complexity as Calabi's diastasis}
\label{C2D}
Let us now relate the circuit complexity of coherent states \cite{Yang,ghmr} 
to the K\"ahler potential.
Complexity measures the degree of difficulty in obtaining
one quantum state from another by means of successive unitary transformations. 
It is thus related to the notion of separation between two states.
There are at least two ways to think about the separation 
between a pair of states.  In one approach the circuit complexity 
between two states is given in terms of their overlap in terms of 
the inner product of states in the corresponding Hilbert space
\cite{takayanagi2018holographic,akal2019reflections}. In the other, 
geometrical, approach, the states are viewed as points on a manifold and the 
separation is the geodesic distance between them. For coherent states 
in here these two notions coincide \cite{ber1,SV,spera1,berceanu1997coherent}.
The overlap of two coherent states given in \eq{xixi} is indeed the 
geodesic distance of points on $\cp^d$,
which, for the Fubini-Study K\"ahler potential \eq{omg:def}, is
given by Calabi's diastasis function \cite{ber1,spera1}.

Calabi's diastasis function is defined for a pair of points 
$(z,z')$ on a K\"ahler manifold as 
\begin{equation}
	\label{calabi:d}
	D(z,z')=\kahler(z,\bar{z})+\kahler(z',\bar{z}')
	-\kahler(z,\bar{z}')-\kahler(z',\bar{z}),
\end{equation}
where 
\begin{equation}
	\label{ozz}
		\kahler(z,z')=\ln (1+\bar{z}\cdot{z'})
\end{equation} 
is obtained from \eq{omg:def}  by analytic continuation. For
small separation of the points it matches with the geodesic distance and
is preserved under restriction to a submanifold \cite{bbdr}.
For the coherent states described above, however, 
the match is exact \cite{ber1}. 
Using \eq{ozz} in \eq{calabi:d} we obtain an expression of the diastasis 
function, which, using \eq{xixi} is related to the overlap of two states as 
\begin{equation}
	\label{Dxi}
	\begin{split}
	D(z,z')&=-2\ln|\braket{z}{z'}|\\
	&= -\ln\ \frac{\big(1+z\cdot \bar z'\big)
	\big(1+\bar z\cdot z'\big)}
	{\big(1+\Vert z\Vert^{2}\big) 
	\big(1+\Vert z'\Vert^{2}\big) }.
	\end{split}
\end{equation}
The geodesic deviation equation with the Fubini-Study metric \eq{FS} is
\begin{equation}
	\label{geo:eq}
	\frac{d^2z_i}{d\tau^2} 
-\sum_{j=1}^d \frac{2\bar{z}_j}{1+\Vert z\Vert^2}\frac{dz_j}{d\tau}
\frac{d z_i}{d\tau} =0,
\end{equation}
where $\tau$ denotes the affine parameter of the geodesic. 
Its solution gives a geodesic as a curve in $\cp^d$ as
\begin{equation}
	\label{geo:sol}
	z_i = \frac{\gamma_i}{\Vert\gamma\Vert}\tan(\Vert\gamma\Vert\tau),
\end{equation}
where $\gamma=(\gamma_i,\gamma_2,\ldots,\gamma_d)$ are complex constants. 
Plugging the solution in  \eq{Dxi} we obtain 
\begin{equation}
	\label{Dtt}
	D(z,z')=-{2}
	\ln\cos\big(\Vert\gamma\Vert(\tau-\tau')\big),
\end{equation}
where $\Vert\gamma\Vert(\tau-\tau')$ is the geodesic distance between the
points $z$ and $z'$ at the values $\tau$ and $\tau'$ of the affine parameter.
The diastasis thus embodies the two definitions of complexity, one in terms
of the overlap of states, or the inner product of sections of the line bundle
$\Ox{\cp^d}{1}$ and the other in terms of the geodesic separation of points. 
The diastasis becomes undefined if the two points
corresponding to the coherent states are separated by $\pi/\Vert\gamma\Vert$.
\section{Deformation of metric}
\label{deform}
The notion of circuit complexity is associated to a cost function. In the 
geometric interpretation of circuit complexity as geodesic distance, the cost
function is given by a change of metric of the manifold of the states. 
Clearly, the cost function can not be completely arbitrary. It must be
compatible with the underlying manifold.  
In particular, the change of the metric is to be such that the topology of 
the space is not 
affected, so that the Chern class in $H^{1,1}(\cp^d)$ is unaltered. 
In this spirit we now consider certain
polarization-preserving deformations of the metric on $\cp^d$ keeping the
curvature fixed \cite{calabi}. 
Let us define
\begin{equation}
\label{s:def}
s = \ln (1+\Vert z\Vert^2).
\end{equation} 
Upon choosing a polarization the K\"ahler potential on $\cp^d$ is a 
function of $s$ in order to be invariant under the circle actions
mentioned before. 
By abuse of notation let us  write  the deformed potential as
\begin{equation}
	\label{K:deform}
\tkahler(z,\bar z)=\tkahler(s).
\end{equation} 
The deformed metric is 
\begin{equation}
\begin{split}
\label{gij}
\tilde{g}_{i\bar\jmath} 
&= \frac{\partial^2\tkahler(z,\bar z)}{\pa z_i\pa\bar z_j}\\
&=e^{-s}\tkahler'(s)\delta_{ij}+e^{-2s}\bar{z}_iz_j\big(\tkahler''(s)-\tkahler'(s)\big),
\end{split}
\end{equation}
where we denote $\tkahler'(s)=\tfrac{\partial\tkahler}{\partial s}$.
Its inverse is
\begin{equation}
\label{inv:g}
\tilde{g}^{i\bar\jmath} = \frac{e^s}{\tkahler'(s)}\delta^{ij}+z_i\bar z_j
\frac{\tkahler'(s)-\tkahler''(s)}{\tkahler'(s)Q'(s)},
\end{equation}
where we define
\begin{gather}
\label{Q}
Q(s)=(1-e^{-s})\tkahler'(s).
\end{gather}
The determinant of the metric tensor is 
\begin{equation}
\label{det:g}
\det\tilde{g} = e^{-sd}(\tkahler'(s))^{d-1}Q'(s).
\end{equation}
In order to obtain this expression we write \eq{gij} in the form of a 
$d\times d$ matrix $I_d + uv^T$ up to an overall
factor, where $I_d$ is the $d\times d$ identity matrix and $u$ and $v$ are two
$d$-dimensional column vectors, $v^T$ being the transpose of $v$. 
Taking the determinant of both sides  of the identity 
\begin{equation}
\begin{pmatrix}
I_d & 0\\ v^T & 1
\end{pmatrix}
\begin{pmatrix}
I_d+uv^T & u\\0 & 1
\end{pmatrix}
\begin{pmatrix}
I_d & 0\\-v^T & 1
\end{pmatrix}
=
\begin{pmatrix}
I_d &  u\\0 & 1+v^T u
\end{pmatrix}
\end{equation}
one derives
\begin{equation}
\det (I_d+uv^T)= 1+v^Tu.
\end{equation}
The expression \eq{det:g} follows from this.
The strategy for finding the deformed K\"ahler potential is to solve for
$\tkahler$  by setting up a differential equation for it 
by first computing the
scalar curvature and then equating it to a constant, customarily taken to 
be $d(d+1)$. However, proceeding to calculate the curvature directly from 
$\tilde{g}_{i\bar{j}}$ will result in a fourth order differential equation for 
$\tkahler$. We resort to an indirect method \cite{calabi}.  Defining 
\begin{gather}
\label{vdg}
v = -\log\det\tilde{g},\\
\psi(s)=\frac{v'(s)}{\tkahler'(s)},
\end{gather}
the Ricci tensor and scalar for the deformed metric assume the forms
\begin{equation}
\begin{split}
R_{i\bar\jmath} &= \frac{\partial^2v}{\pa z_i\pa\bar z_j}\\
&= 
e^{-s}v'(s)+e^{-2s}\bar{z}_iz_j\big(v''(s)-v'(s)\big),
\end{split}
\end{equation}
and 
\begin{equation}
\begin{split}
\label{R}
R(s)&=g^{i\bar\jmath}R_{i\bar\jmath}\\
&= d\psi(s)+\frac{Q(s)}{Q'(s)}\psi'(s),
\end{split}
\end{equation}
respectively. 
Setting the Ricci scalar of $\cp^d$ to the constant,
$R(s)=d(d+1)$, as is customary, and using \eq{Q} in \eq{R}
we solve for $\psi$ as
$\psi(s) = d+1$, that is,
\begin{equation}
v'(s)=(d+1)\tkahler'(s).
\end{equation}
Using \eq{vdg} this leads to
\begin{equation}
\label{detg:sol}
\det\tilde{g} = g_0\ e^{-(d+1)\tkahler},
\end{equation}
where  $g_0$ is a constant of integration.
We now have two expressions for the volume factor $\det\tilde{g}$, namely, 
\eq{det:g}, obtained from the definition of the metric in terms of the K\"ahler 
potential, and \eq{detg:sol}, obtained by solving the equation for 
the constancy of scalar curvature. 
Equating these two expressions  we obtain 
\begin{equation}
\label{detg2}
g_0\ e^{-(d+1)\tkahler} =
e^{-sd}(\tkahler'(s))^{d-1}Q'(s).
\end{equation}
Differentiating with respect to $s$ and using this once again along with
\eq{Q} to eliminate $\tkahler'$ we obtain 
\begin{equation}
\label{eq:Q-d}
\frac{d}{ds}\left(\frac{e^{-sd}Q(s)^{d-1}}{(1-e^{-s})^{d-1}} Q'(s)\right)
+(d+1) \frac{e^{-sd}Q(s)^d}{(1-e^{-s})^d}Q'(s)=0.
\end{equation}
Once the function $Q(s)$ is obtained by solving this differential equation,
the K\"ahler potential can be evaluated either as an integral of $Q$ using 
\eq{Q}, or as a derivative of $Q$ using \eq{detg2} in conjunction with
\eq{Q}. From the latter we obtain
\begin{equation}
\label{KfromQ}
\tkahler(s) = -\frac{1}{d+1}\ln\left(e^{-sd}\left(
\frac{Q(s)}{1-e^{-s}}\right)^{d-1}Q'(s)\right).
\end{equation}

These can be recast in terms of 
$X=\sum_{i=1}^d x$, 
changing variable from $s$ to $X$, as
$s=\ln (1+X)$. The differential equation \eq{eq:Q-d} assumes the form
\begin{equation}
\label{Q:X}
\frac{d}{dX}\left(\left(\frac{Q}{X}\right)^{d-1}\frac{dQ}{dX}\right)
+ (d+1) \left(\frac{Q}{X}\right)^d\frac{dQ}{dX}=0,
\end{equation}
where now $Q$ is taken to be a function of $X$.
Moreover, 
The K\"ahler potential \eq{K:deform} can be looked upon as a deformation 
of \eq{F:pn} as
\begin{equation}
	\label{FK}
	\tF(X) = \tkahler(s),
\end{equation}
where, by \eq{Q}, 
\begin{equation}
	\label{QF}
	Q(X) = X\frac{d\tF(X)}{dX}.
\end{equation}
Equation \eq{Q:X} can be further cast into
\begin{equation}
\label{QX:gen}
X Q \frac{d^2Q}{dX^2} + (d+1) Q^2\frac{dQ}{dX}
+(d-1)\left(X\frac{dQ}{dX}-Q\right)\frac{dQ}{dX}=0 .
\end{equation} 
In the light of \eq{FK}, the expression \eq{KfromQ} takes the form 
\begin{equation}
\label{FfromQ}
\tF(X) = -\frac{1}{d+1}\ln \left(\left(\frac{Q}{X}\right)^{d-1}
\frac{dQ}{dX}\right).
\end{equation}
\begin{example}
\label{Deform1}
The equation \eq{FfromQ} is non-linear, not giving in to analytic methods of
solution except for the simple case of $d=1$.
In this case  $X=x$, since $x$ now has a single component.
The metric is 
\begin{equation}
	\label{gQ}
	\tilde{g}_{z\bar{z}} = Q(x)',
\end{equation} 
where a prime is taken to denote a derivative with respect to $x$, \emph{i.e.,}
${}^{\prime}=\tfrac{d}{dx}$ for this example. We have 
\begin{equation}
\label{vdg1}
	v = -\ln\det\ \tilde{g}=-\ln \tilde{g}_{z\bar{z}}, 
\end{equation} 
the Ricci tensor and the scalar curvature are, respectively,
\begin{gather}
	R_{z\bar{z}} = \partial\bar{\partial}v
		=(xv')',\\
\label{R1}
	R(s)=\tilde{g}^{z\bar{z}}R_{z\bar{z}}
	=(xv')'/Q'(x).
\end{gather} 
Setting $R(s)=2$ we have
\begin{equation}
	(xv'-2Q(x))'=\big(x(v-2\tF(x))'\big)'=0.
\end{equation}
One solution of this is $v=2\tF(x)$, which, using \eq{vdg1}  and \eq{gQ} yields
\begin{equation}
\label{K:vol}
	Q'(x) = e^{-2\tF(x)},
\end{equation} 
where we fixed a constant of integration thereby fixing the normalization 
of volume. 
Differentiating once again we obtain a second order equation for $Q$ as
\begin{equation}
	\label{eq:Q-d1}
	xQ''+{2}QQ'=0,
\end{equation}
same as \eq{QX:gen} with $d=1$.
The general solution to this non-linear equation is 
\begin{equation}
\label{Q1dim}
	Q(x)= \frac{1}{2}-\frac{\alpha}{2}\left(
	\frac{1-\beta x^{\alpha}}{1+\beta x^{\alpha}}\right),
\end{equation} 
where $\alpha$ and $\beta$  are constants. 
From \eq{FfromQ}
\begin{gather}
\label{kah:disF}
	\tilde{\F}(x)=-\frac{1}{2} \ln Q'(x) = -\frac{1}{2}
\ln\frac{\alpha^2\beta x^{\alpha-1}}{(1+\beta x^{\alpha})^2},\\
\label{kah:disK}
\tkahler(z,\bar{z}) = -\frac{1}{2}
\ln\frac{\alpha^2\beta |z|^{2(\alpha-1)}}{(1+\beta |z|^{2\alpha})^2},
\end{gather} 
using the relation \eq{zeta}.
The numerator inside the logarithm in \eq{kah:disK} gives rise to
additive holomorphic and anti-holomorphic terms, $\ln z$ and $\ln\bar{z}$, 
respectively. They do not affect the K\"ahler form or the metric. 
The deformed metric is
\begin{equation}
\label{g:new}
		\tilde{g}_{z\bar{z}} = 
		\frac{\alpha^2\beta |z|^{2(\alpha-1)}}{(1+\beta |z|^{2\alpha})^2},
\end{equation}
which reduces to the Fubini-Study metric \eq{FS} for $\alpha=\beta=1$.
However, since we associate the K\"ahler potential itself
with entropy and complexity, we need to retain these factors depending 
on the constants. The coherent states \eq{z:eq} need to be normalized 
anew with \eq{omg:def} generalized
to \eq{kah:disK}. However, it can be checked by explicitly
writing the diastasis function \eq{calabi:d} with \eq{kah:disK} and the 
new coherent states that the equality in the first line of \eq{Dxi} holds
no more. The Fubini Study K\"ahler potential and some deformed ones are
shown in Figure~\ref{fig:d1}.
\end{example}

For $d>1$ we resort to solving \eq{QX:gen} numerically. Setting $Q(0)=0$
and $\frac{dQ(0)}{dX}=1$ yields the Fubini Study potential in \eq{FfromQ}.
Deformations correspond to different boundary conditions for \eq{QX:gen}. Some 
such instances are shown in Figure~\ref{fig:kah} for different dimensions.
From these plots is appears that the deformations are similar in form
in  all dimensions, cf. \cite{Shan}. Hence, the failure of the \eq{Dxi} 
is expected to continue across dimension.

\begin{figure}[h]
\begin{subfigure}[b]{.52\textwidth}
\centering
\begin{tikzpicture}
\pgfplotstableread{Fdim1.dat}
\dimone
\begin{axis}[
xmin=0,xmax=30,
ymin=0,ymax=5,
x label style={at={(axis description cs:.7,-0.07)},anchor=north},
y label style={at={(axis description cs:-0.13,.89)},anchor=north},
xlabel=$\scriptstyle X$,
ylabel=$\scriptstyle \tF(X)$,
legend style = { at = {(1,.3)}}]
\addplot[mark=] table[y = a] from \dimone ;
\addlegendentry{\tiny Fubini Study $Q'(0)=.25$}
\addplot[dashed,mark=] table[y = b] from \dimone ;
\addlegendentry{\tiny deformed $Q'(\epsilon)=10$}
\addplot[thick,dotted,mark=] table[y = c] from \dimone ;
\addlegendentry{\tiny deformed $Q'(\epsilon)=0.1$}
\end{axis}
\end{tikzpicture}
\caption{$d=1$}
\label{fig:d1}
\end{subfigure}
\begin{subfigure}[b]{.5\textwidth}
\centering
\begin{tikzpicture}
\pgfplotstableread{Fdim3.dat}
\dimthree
\begin{axis}[
xmin=0,xmax=30,
ymin=0,ymax=5,
x label style={at={(axis description cs:.7,-0.07)},anchor=north},
y label style={at={(axis description cs:-0.13,.89)},anchor=north},
xlabel=$\scriptstyle X$,
ylabel=$\scriptstyle \tF(X)$,
legend style = { at = {(1,.3)}}]
\addplot[mark=] table[y = a] from \dimthree ;
\addlegendentry{\tiny Fubini Study $Q'(0)=.25$}
\addplot[dashed,mark=] table[y = b] from \dimthree ;
\addlegendentry{\tiny deformed $Q'(\epsilon)=10$}
\addplot[thick, dotted,mark=] table[y = c] from \dimthree ;
\addlegendentry{\tiny deformed $Q'(\epsilon)=0.1$}
\end{axis}
\end{tikzpicture}
\caption{$d=3$}
\label{fig:d3}
\end{subfigure}
\begin{subfigure}[b]{.52\textwidth}
\centering
\begin{tikzpicture}
\pgfplotstableread{Fdim7.dat}
\dimseven
\begin{axis}[
xmin=0,xmax=30,
ymin=0,ymax=5,
x label style={at={(axis description cs:.7,-0.07)},anchor=north},
y label style={at={(axis description cs:-0.13,.89)},anchor=north},
xlabel=$\scriptstyle X$,
ylabel=$\scriptstyle \tF(X)$,
legend style = { at = {(1,.3)}}]
\addplot[mark=] table[y = a] from \dimseven ;
\addlegendentry{\tiny Fubini Study $Q'(0)=.25$}
\addplot[dashed,mark=] table[y = b] from \dimseven ;
\addlegendentry{\tiny deformed $Q'(\epsilon)=10$}
\addplot[thick,dotted,mark=] table[y = c] from \dimseven ;
\addlegendentry{\tiny deformed $Q'(\epsilon)=0.1$}
\end{axis}
\end{tikzpicture}
\caption{$d=7$}
\label{fig:d7}
\end{subfigure}
\begin{subfigure}[b]{.5\textwidth}
\centering
\begin{tikzpicture}
\pgfplotstableread{Fdim20.dat}
\dimtwnty
\begin{axis}[
x label style={at={(axis description cs:.7,-0.07)},anchor=north},
y label style={at={(axis description cs:-0.13,.89)},anchor=north},
xlabel=$\scriptstyle X$,ylabel=$\scriptstyle\tF(X)$,
xmin=0,xmax=30,
ymin=0,ymax=5,
legend style = { at = {(1,.3)}}]
\addplot[mark=] table[y = a] from \dimtwnty ;
\addlegendentry{\tiny Fubini Study $Q'(0)=.25$}
\addplot[dashed,mark=] table[y = b] from \dimtwnty ;
\addlegendentry{\tiny deformed $Q'(\epsilon)=10$}
\addplot[thick,dotted,mark=] table[y = c] from \dimtwnty ;
\addlegendentry{\tiny deformed $Q'(\epsilon)=0.1$}
\end{axis}
\end{tikzpicture}
\caption{$d=20$}
\label{fig:d20}
\end{subfigure}
\caption{K\"ahler potential in various dimensions for boundary conditions
$Q(\epsilon)=\epsilon$ and $Q'(\epsilon)$ as indicated in the plots,
with $\epsilon=10^{-5}$.}
\label{fig:kah}
\end{figure}

\section{Conclusion}
In this article
we point out a connection between entropy and complexity
of coherent states of $SL(d+1,\C)$. Both can be written in terms of the 
Fubini-Study K\"ahler potential on $\cp^d$. The coherent states
are interpreted as global sections of a spin $\tfrac{1}{2}$ 
bundle on $\cp^d$, with 
an inner product defined using the first chern class. 
Writing the Legendre transform of the 
K\"ahler potential in terms of the dual symplectic variable the resulting 
Guillemin potential is shown to be the entropy for the coherent states.
While Wehrl entropy \cite{weh78} is defined for other spin states, it
can not be related to the Guillemin potential. 
Complexity is identified with Calabi's diastasis function, which is 
a combination of the analytic continuation of the K\"ahler potential as well.
Moreover, we indicate that these identifications fail for otherwise
reasonable deformations of the K\"ahler potential. For a quantum system
the notion of complexity is the lack of ease for the system to evolve from
one state to another. It is fixed by optimising with respect to allowed 
parameters of the system, like the integration constants appearing in
solving \eq{FfromQ}. A variational function of these, called the cost 
function \cite{ghmr} can be used for this. From the analysis here 
the Fubini-Study potential appears to be the one of optimal cost. 
%
\section*{Acknowledgement}
I thank Rohit Mishra and E. Gasparim for useful interactions.

\end{document}